\documentclass[screen,sigplan,nonacm]{acmart}

\setcopyright{none}
\renewcommand\footnotetextcopyrightpermission[1]{}

\usepackage{amsmath}
\usepackage{algpseudocode}
\usepackage{graphicx}
\usepackage{subcaption}
\usepackage{multirow}
\usepackage{diagbox}
\usepackage{threeparttable}
\usepackage{booktabs}
\usepackage{makecell}
\usepackage{pifont}
\usepackage{enumitem}
\usepackage{bbding}
\usepackage[ruled,vlined]{algorithm2e}
\SetKwProg{Fn}{Function}{:}{end}

\AtBeginDocument{%
  \providecommand\BibTeX{{%
    \normalfont B\kern-0.5em{\scshape i\kern-0.25em b}\kern-0.8em\TeX}}}

\begin{document}

\newcommand\blfootnote[1]{%
\begingroup
\renewcommand\thefootnote{}\footnote{#1}%
\addtocounter{footnote}{-1}%
\endgroup
}
\newcommand{\p}[1]{\noindent\textbf{#1}\hspace{0.5em}\ignorespaces}
\title{ Physically Partitioned KVCache Format for CPU--GPU Load Balancing in MoE Inference}
\author{Enda Yu, Dezun Dong\textsuperscript{*}, Xiangke Liao}
\email{{yuenda,dong,xkliao}@nudt.edu.cn}
\affiliation{%
  \institution{National University of Defense Technology}
  \country{China}
}




\begin{abstract}
Single-GPU long-context inference with Mixture-of-Experts (MoE) models requires spilling the key-value cache (KVCache) to CPU memory.
The spilled KV serves two complementary purposes---transferring to the GPU for attention computation, or computing in-place on the CPU---which demand opposing physical states.
The optimal split between them varies with workload, yet existing KVCache abstractions offer only storage semantics over a monolithic object of a single physical state, and cannot express dynamic load balancing.
We propose InplaceKVCache, the first KVCache abstraction whose format fixes each byte's physical residency at write time, so that the CPU--GPU load balance can be adjusted without moving data after placement.
It realizes this as a four-region layout along two dimensions---device affinity and access pattern---turning load balancing into pure scheduling.
Built on this abstraction, WriteScope splits CPU--GPU shares along the sequence dimension, and a portable roofline performance model determines the optimal CPU share as sequence length evolves, with online feedback tracking CPU cost drift.
On three MoE models (DeepSeek-V2-Lite, Qwen3-30B-A3B, Mixtral-8$\times$7B) with a 32~GB VRAM budget, WriteScope supports end-to-end inference at the 1M-token aggregate scale.
In the long-context regime ($\ge$8K), it achieves geometric-mean speedups of $1.5\times$--$2.5\times$ on A100 and $1.4\times$--$1.7\times$ on V100 over four reproduced baselines, while vLLM, SGLang, and KTransformers fail even with a doubled KV budget.
A DeepSeek-V4-Flash case study validates composition with native sparse attention.\blfootnote{*Dezun Dong is the Corresponding Author}
\end{abstract}

\maketitle

\begin{figure}[t]
  \centering
  \begin{subfigure}[ht]{\linewidth}
    \includegraphics[width=\linewidth]{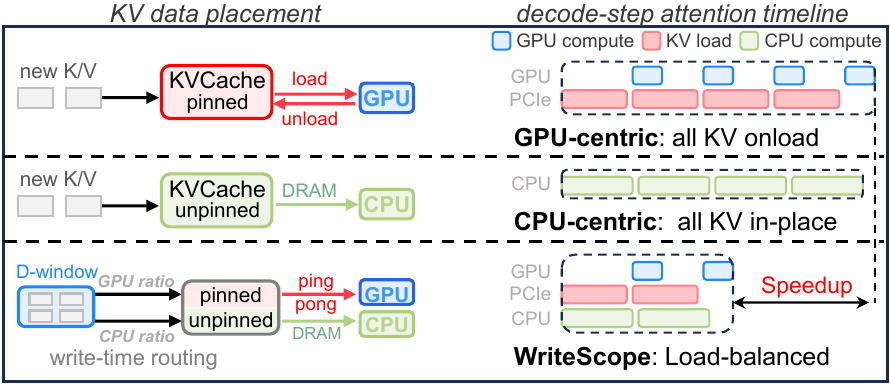}
  \end{subfigure}
  \caption{Three KVCache placement strategies. WriteScope routes KV to fixed physical partitions at write time, enabling dynamic CPU-GPU load balancing without reorganization.}
     \label{figure1} 
\end{figure}

\section{Introduction}
\label{sec:intro}

By virtue of sparse activation---each token activates only a handful of experts~\cite{shazeer2017moe,switchtransformer}---MoE architectures are a natural fit for budget-constrained, privacy-first deployment on a single GPU~\cite{flashllm,PowerInfer}, whether a personal machine or an on-premise server.
In this setting, single-user long-context inference is emerging as a core demand: per-turn inputs stay modest while the accumulated context grows relentlessly, driven by long-horizon agent reasoning, persistent-memory conversations, and parallel rollouts.
Yet the sum of model weights and long-context KVCache far exceeds what a single GPU in this setting can hold: for Qwen3-30B-A3B~\cite{Qwen3}, inference with an aggregate 128K-token context requires at least 78~GB, well beyond the 32~GB of a single V100.
The KVCache, together with most weights, must spill to CPU-resident memory.

The spilled KV has two complementary uses, and existing CPU-GPU hybrid inference systems adopt one or the other.
GPU-centric approaches load all KV back to the GPU (e.g., FlexGen~\cite{flexgen}, SolidAttention~\cite{SolidAttention}); CPU-centric approaches compute all KV in-place on the CPU (e.g., NEO~\cite{neo}, ScoutAttention~\cite{scoutattention}).
In both cases the cost---PCIe transfer for the former, CPU compute for the latter---scales with the full KV volume, so a single strategy inevitably hits a bottleneck on one side as the workload grows.
In long-context regimes, both camps mitigate this with asynchronous pre-execution of the next layer's attention during the FFN phase (\emph{AF overlap})~\cite{InfiniGen}.
For small-scale dense models, the FFN is executed entirely on the GPU, leaving the CPU and PCIe idle during that phase and available for attention \emph{opportunistically}.

Yet MoE ends the free lunch: spilled expert parameters reside on the CPU, so expert computation and loading saturate both CPU and PCIe during the FFN phase, turning that overlap into contention.
On GPU-centric SolidAttention (Qwen3-30B-A3B, batch size 16, sequence 32K), pre-loading the next layer's KVCache during FFN raises attention latency from 33.7~ms to 74.2~ms and drops KV transfer bandwidth from 25.4~GB/s to 14.9~GB/s.
Once overlap breaks down, the PCIe-bandwidth bottleneck of the GPU-centric camp and the CPU-compute bottleneck of the CPU-centric camp stand fully exposed, and the only lever left is the volume each engine must handle.
Moreover, no fixed division of that volume suffices: the CPU's effective per-token cost shifts with working set and cache locality while the GPU's is set by PCIe transfer, so the balance point drifts with model, batch, and sequence length (\S\ref{sec:load-balance-need}).
\emph{Dynamic load balancing} of the KVCache is therefore not an optimization but a requirement.

However, existing KVCache abstractions cannot express this balance: they promise only storage semantics (store, retrieve) over a monolithic object of one physical state---which bytes go to the CPU and which to the GPU is outside their vocabulary.
Worse, the two uses impose opposite physical-state requirements: data destined for GPU loading must be pinned and logically contiguous (otherwise DMA bandwidth collapses), whereas data destined for CPU computation must be unpinned and organized per head (otherwise CPU reads are penalized).
Implementations that try to serve both must allocate auxiliary contiguous buffers, copy, and reorganize---preparation that balloons from 3.5~ms to 185.9~ms per layer per step as context grows from 2K to 64K (\S\ref{sec:timing})---negating the gains of load balancing.
This is why the two camps each stick to one extreme.
The difficulty of dynamic load balancing lies not in scheduling, but in the data structure.

Our key insight is a data-layout principle we call the \emph{write-time principle}: a byte has exactly one final physical residency, and that residency is determined at write time.
Physical ownership thus becomes a property of the data format rather than a runtime mechanism, and load balancing needs no post-hoc reorganization.
Fig.~\ref{figure1} contrasts the resulting placement with both extremes: KV routed to fixed physical partitions at write time, balancing smoothly with load.
This principle fits the KVCache because its consumption is deterministic---unlike stochastic expert activation, which can only be arbitrated at schedule time (e.g., Fiddler~\cite{fiddler}).

Building on this insight we propose \emph{InplaceKVCache}, the cache format, and \emph{WriteScope}, the scheduling layer built on it.
InplaceKVCache partitions space along two orthogonal dimensions, device affinity and access pattern, into four regions: an unpinned compute region and a pinned load region on the CPU side, and a ping-pong compute region and a D-window offload region on the GPU side.
Each engine reads only its own region---the CPU compute engine works over the unpinned compute region in place, the GPU load engine streams the pinned load region through ping-pong---so the entire fallback path of auxiliary buffers, copies, and reorganization disappears.
Prefill KV is routed to its region at write time; decode KV is staged in the D-window and, once an eviction batch of $D$ entries fills, routed whole to the pinned or unpinned region. After placement nothing moves again.
On top of this abstraction, WriteScope splits CPU/GPU shares along the sequence dimension. A portable roofline model directs each eviction batch to the region keeping the cumulative share on the roofline target, equalizing the GPU's transfer-plus-compute cost against the CPU's compute cost.

On three MoE models spanning both MLA and GQA~\cite{gqa} attention architectures (DeepSeek-V2-Lite~\cite{deepseekv2}, Qwen3-30B-A3B~\cite{Qwen3}, Mixtral-8$\times$7B~\cite{mixtral}), with a 32~GB VRAM budget, WriteScope supports end-to-end inference at the 1M-token scale (bs=32 $\times$ seq=32K).
In the long-context regime ($\geq$8K), it achieves geometric-mean speedups of 1.5$\times$--2.5$\times$ on A100 and 1.4$\times$--1.7$\times$ on V100 over four reproduced baselines, while vLLM~\cite{vllm}, SGLang~\cite{sglang}, KTransformers~\cite{ktransformers} fail broadly.
A DeepSeek-V4~\cite{ds4} case study validates WriteScope's generalization to native sparse-attention scenarios. Contributions:
 
\begin{enumerate}[leftmargin=*]
    \item \textbf{Problem formulation and root-cause attribution} (\S\ref{sec:background}--\ref{sec:why-monolithic}).
    We define the core requirement of CPU-GPU hybrid attention as dynamic load balancing of the KVCache, trace its long-standing absence to the monolithic storage semantics of existing KVCache abstractions, and quantitatively reveal the systematic breakdown of AF overlap under MoE inference on resource-constrained platforms.
\item \textbf{The InplaceKVCache abstraction} (\S\ref{sec:inplace}).
The first KVCache abstraction whose format fixes physical residency (device affinity $\times$ access pattern) at write time; its orthogonal four-region layout makes dynamic load balancing a built-in capability---an online roofline policy retunes the CPU/GPU share while placed data never moves.
\item \textbf{Load-balancing scheduling} (\S\ref{sec:scheduling}). A sequence-dimension split of CPU/GPU shares, grounded in a roofline model that tracks the load-balance point as it drifts with model, batch size, and sequence length.    
\end{enumerate}

The vision WriteScope points to is this: long-context MoE inference no longer requires a multi-GPU server---it becomes a practical configuration on the single-GPU machines that privacy and compliance considerations already favor.


\section{Background and Requirements Analysis}
\label{sec:background}

\begin{table}[t]
\centering
\caption{VRAM breakdown (bs=4 $\times$ seq=32K, bf16, GB).}
\label{tab:memory}
\resizebox{\linewidth}{!}{
\begin{tabular}{lccccc}
\toprule
\textbf{Model} & \textbf{Weights} & \textbf{KV} & \textbf{CUDA} & \textbf{Temp} & \textbf{Total} \\
\midrule
DeepSeek-V2-Lite  & 31.4 & 4.1 & 2.4 & 1.2 & 39.1 \\
Qwen3-30B-A3B  & 61.1 & 12.9 & 2.4 & 1.8 & 78.2 \\
Mixtral-8$\times$7B  & 93.4 & 17.2 & 2.4 & 1.7 & 114.7 \\
\bottomrule
\end{tabular}
}
\footnotesize
\textbf{Temp}: peak runtime allocation (activations and intermediates).
\end{table}

\subsection{KV Is the Right Object to Spill}
\label{sec:spill-object}

Table~\ref{tab:memory} quantifies the overflow: for three models at an aggregate 128K tokens, the measured VRAM demand---including CUDA context and temporary buffers---reaches 1.2$\times$, 2.4$\times$, and 3.6$\times$ a 32~GB card, so spilling is unavoidable.
All measurements in this paper accordingly cap the A100-40GB allocation at 32~GB---matching the V100-32GB used for portability validation, so both platforms run under one budget.

Which object should be spilled---weights or KV?
On Qwen3 (bs=4, seq=2K), raising vLLM's KV budget from 2~GB to 11~GB and pushing more expert weights to CPU increases prefill latency by 53\% and decode latency by 51\%.
Expert GEMM has far higher arithmetic intensity than attention, so keeping weights GPU-resident yields greater benefit~\cite{fiddler,layerscope}.
KV, not weights, is therefore the right object to spill.

The next question is when to evict.
Existing frameworks evict reactively: the vLLM swap engine~\cite{vllm} evicts old blocks only when the GPU block pool is exhausted; SGLang's HiCache~\cite{sglang} writes back to host only on radix hit or eviction.
Under a 4~GB KV budget, vLLM slows significantly once block-pool demand reaches 1.8$\times$ capacity and deadlocks at 2.9$\times$ once the KV injection rate exceeds the eviction rate; SGLang fails allocation at merely 1.08$\times$.
DRAM capacity is not the bottleneck; eviction timing is.
Eviction decisions must therefore be made proactively, at write time.

\subsection{The Need for KVCache Load Balancing}
\label{sec:load-balance-need}

CPU-resident KVCache has two complementary uses: \emph{loading} (required KV transferred over PCIe to GPU for attention, bounded by PCIe bandwidth) and \emph{in-place computation} (KV stays in DRAM and attention is executed by CPU cores, bounded by CPU compute).
Each decode step can load a portion of the KV while computing the rest in-place, with both engines working concurrently.
The optimal CPU share varies widely with workload: for Qwen3-30B-A3B it spans 0 to 0.625 (i.e., 62.5\% of KV computed in-place, \S\ref{sec:roofline}).
No fixed share covers this spectrum, so an efficient system must select the proportion of the two uses for its (model, batch, sequence length) and re-select it as the sequence grows.

\begin{figure}[t]
  \centering
  \begin{subfigure}[ht]{\linewidth}
    \includegraphics[width=\linewidth]{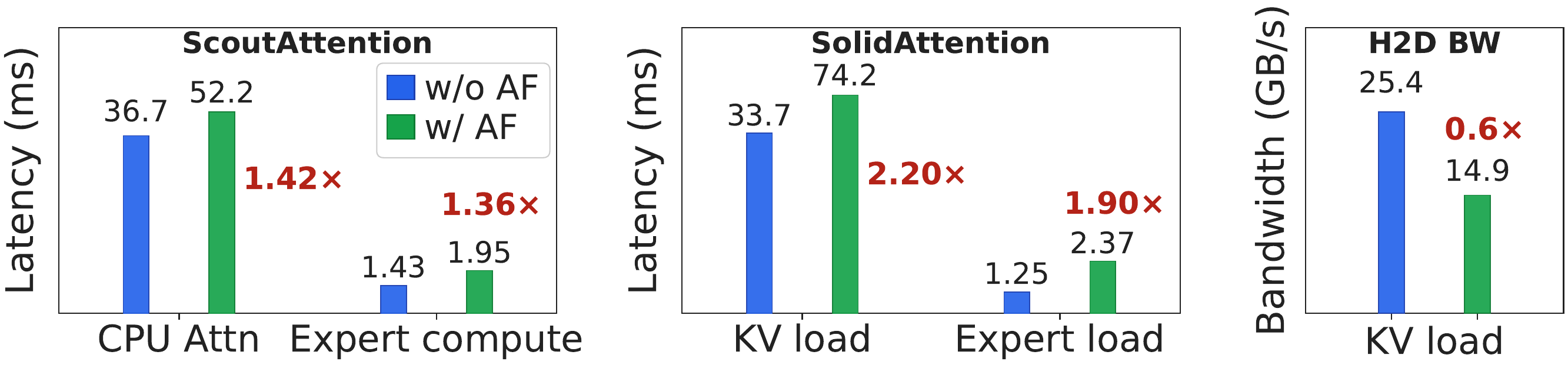}
  \end{subfigure}
  \caption{AF overlap contention under MoE (Qwen3-30B-A3B, bs=16, seq=32K). AF degrades attention, expert, and PCIe bandwidth simultaneously due to resource contention.}
     \label{figure2} 
\end{figure}

\subsection{AF Overlap Breaks Under MoE}
\label{sec:af-break}

Lacking dynamic load balancing, existing systems commit exclusively to one use: GPU-centric approaches rely on \emph{loading}, CPU-centric ones on \emph{in-place computation}.
To mitigate their respective costs, both camps employ AF overlap---GPU-centric systems via pre-loading (early KV transfer, lossless), CPU-centric systems via pre-computation (predicting the next layer's attention from adjacent-layer input similarity, at the cost of accuracy).
This remedy relies on a premise: the FFN executes entirely on the GPU, leaving the CPU and PCIe completely idle during that phase (true for dense models).

MoE invalidates this premise: weight spilling leaves most expert parameters CPU-resident.
Expert placement follows the greedy schedule common to MoE offloading systems~\cite{layerscope,hybridmoe}: during FFN, the CPU executes cold-expert GEMM locally while the PCIe is busy loading hot-expert weights to the GPU---precisely the resources AF overlap set out to borrow.
Fig.~\ref{figure2} quantifies the resulting contention on two representative implementations (SolidAttention and ScoutAttention; Qwen3-30B-A3B, bs=16, seq=32K; AF-enabled vs.\ AF-disabled): AF degrades the attention path by 1.4$\times$--2.2$\times$, expert-side loading/compute by 1.4$\times$--1.9$\times$, and KV host-to-device (H2D) bandwidth to approximately 0.6$\times$.
AF overlap thus turns from an enabler into a net penalty: the pre-execution meant to hide inside the FFN phase now contends with expert work for the same saturated CPU and PCIe.



\section{Why Monolithic Caches Fail to Balance}
\label{sec:why-monolithic}

The write-time principle (\S\ref{sec:intro}) fixes a byte's physical residency---its pin state and layout contiguity, two orthogonal attributes---once, at write time.
This section grounds the principle in three microbenchmarks---two on the attributes (\S\ref{sec:pinning}, \S\ref{sec:strided}) and one on decision timing (\S\ref{sec:timing})---which together show why monolithic abstractions cannot balance efficiently (\S\ref{sec:format-conclusion}).
All measurements use DS2 (DeepSeek-V2-Lite~\cite{deepseekv2}) on the A100 platform (2 warmup iterations, mean of 10 runs, run-to-run variance $<$2\%).

\subsection{PCIe Transfer and Compute Differ on Pinning}
\label{sec:pinning}

PCIe transfer and in-place computation impose opposite pin requirements: DMA loading to the GPU requires pinned memory~\cite{cudapinned}, otherwise falling back to a low-bandwidth synchronous path, whereas the CPU reads pinned memory more slowly than unpinned (page-table and allocation-path penalties).
Measured on a single-layer KVCache buffer (sweeping seq 2K$\rightarrow$32K), pinned transfer is consistently 2.5$\times$ faster (25.4 vs.\ 10~GB/s), while CPU reads of pinned run 10--25\% slower beyond 8K.
Pin state is thus a \emph{seesaw}: whichever state a monolithic cache commits to favors one engine and penalizes the other---serving different parts of the same cache with different engines is unrealizable under a single pin state.

\begin{figure}[t]
  \centering
  \begin{subfigure}[ht]{\linewidth}
    \includegraphics[width=\linewidth]{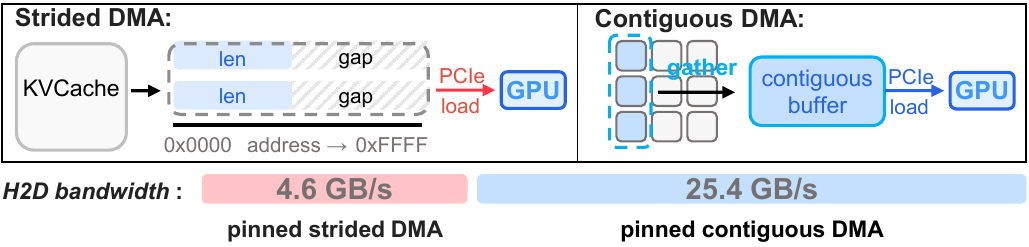}
  \end{subfigure}
\caption{Strided vs.\ contiguous DMA on a pre-allocated buffer. Inter-row gaps fragment the source into short segments, forcing per-segment copies; a contiguous layout admits one full-bandwidth transfer.}
     \label{figure3} 
\end{figure}

\subsection{Strided DMA Collapses H2D Bandwidth}
\label{sec:strided}

Layout contiguity is the second attribute---even with correct pinning, a single full-bandwidth DMA requires the source to be contiguous in virtual address space.
To avoid frequent allocation and copying, frameworks typically pre-allocate a single maximum-capacity buffer for reuse~\cite{huggingface}.
Fig.~\ref{figure3} illustrates the consequence: when the actual sequence length falls short of capacity, each row of valid data is followed by a gap of unused capacity, fragmenting the slice into short contiguous segments.
DMA cannot merge such a source into one large transfer and must copy segment by segment synchronously, with per-segment overhead dominating.

In a 32K pre-allocated buffer (sequence 2K$\rightarrow$32K), strided slices achieve only 4.6~GB/s, recovering to 25.4~GB/s only at the full 32K where the gap vanishes; exactly-allocated contiguous buffers hold 25.4~GB/s throughout---a 5.5$\times$ shortfall.
Pre-allocation thus trades contiguity for allocation-free reuse---a variable-length workload cannot keep a fixed buffer exactly allocated, so the gap is the common case.

\subsection{State Is Freely Chosen Only at Write Time}
\label{sec:timing}

The cost of changing physical state (re-pinning, re-layout) is proportional to data volume, so timing decides the total cost. 
At write time the data is already in flight: choosing its pin state and layout adds no copy traffic, and the pin work itself is paid once, on new bytes only, where it can be scheduled ahead of use (\S\ref{sec:interface}).
Afterwards, the same choice costs a full copy of everything already placed---on the critical path, because the consumer is already waiting.
This is what makes the pinning and contiguity asymmetries binding---a monolithic cache caught on the wrong side of either can escape only by paying that copy.

Existing implementations fare even worse: they do not change state once---they rebuild the buffer every step.
vLLM's blocks fix pin state and layout at allocation time, so post-hoc migration re-walks the full prepare pipeline of loading$\rightarrow$ splicing$\rightarrow$pinning$\rightarrow$chunked H2D.
On DynamicCache (bs=16, all KV consumed on GPU), this prepare pipeline costs 3.5~ms per layer per step at a 2K context, growing superlinearly to 185.9~ms at 64K---an order of magnitude above the layer's entire attention compute.
Physical state must therefore be set at write time, as a single atomic operation with the write.

\subsection{Physical Ownership Must Be in the Format}
\label{sec:format-conclusion}

The three measurements together show that monolithic abstractions struggle to support CPU-GPU load balancing efficiently on our target platform.
A surface-level workaround is dual copies---maintaining two physical-state copies of every byte, one for each engine, with dynamic ratio and zero reorganization.
Dual copies are infeasible: first, no reusable format---monolithic abstractions lack an existing contiguous pinned block format that can be sliced per chunk, so the GPU-side copy requires an extra preprocessing step to copy and convert spilled data into independent contiguous blocks; second, double capacity---dual-copy KVCache for Mixtral bs=32$\times$seq=32K reaches 274~GB, enough alone to exhaust the host memory of 256~GB-class or smaller devices.
Nor can the problem be lifted into a runtime layer: pin state and layout are properties of the bytes themselves, so a wrapper above a single-state pool inherits its seesaw and stride penalties unchanged.\footnote{These costs concern reorganizing already-placed data; the D-window's batched first placement of new KVCache (\S\ref{sec:routing}) touches none.}

The design corollary is thus direct: physical ownership

\noindent belongs in the data format, not in a runtime mechanism.
\section{The InplaceKVCache Format}
\label{sec:inplace}

\begin{figure}[t]
  \centering
  \begin{subfigure}[ht]{\linewidth}
    \includegraphics[width=\linewidth]{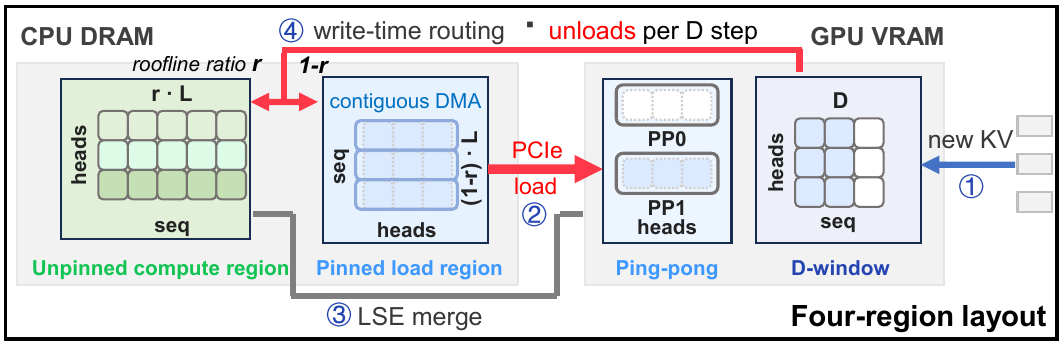}
  \end{subfigure}
\caption{InplaceKVCache four-region layout. The roofline-guided share $r$ steers write-time routing of eviction batches from the GPU D-window to CPU unpinned or pinned regions. The CPU computes the unpinned share; the GPU loads the pinned share into ping-pong, with outputs merged via LSE.}
     \label{figure4} 
\end{figure}

\subsection{The Format Is the Partition}
\label{sec:format}

InplaceKVCache refactors the KVCache from a monolithic object with a single physical state into a data format with explicit physical partitioning: a four-region layout along two orthogonal dimensions, device affinity and access pattern (Fig.~\ref{figure4}).
CPU$\times$compute maps to the unpinned compute region, CPU$\times$transfer to the pinned load region, GPU$\times$transfer to the D-window offload region, and GPU$\times$compute to the ping-pong compute region.
Each region is internally organized in fixed-size blocks; the block is the unified granularity for write-time routing, staging, eviction, and transfer.
Write-time routing thus places each K/V entry into a fixed region, and InplaceKVCache is constructed without the intermediate monolithic state that would require post-hoc slicing.
As \S\ref{sec:intro} argues, KVCache consumption is deterministic, so the pin$\times$layout duality can be settled in the format.

\subsection{Batched Write-Time Routing in Decode}
\label{sec:routing}

Prefill and decode route differently; the crux is when the written KV is next consumed.
Prefill KV is generated in bulk and can be routed to its final destination in one shot at write time; decode tokens must be attended starting from the next step, so their destination can wait one eviction window, and we therefore route in batches.\footnote{Whole-prompt prefill is balancing-free: each layer's KV (151--537~MB at bs=4$\times$seq=32K) is consumed by attention the moment it is produced and routed away on the idle D2H direction.}

New decode K/V is first staged in the GPU's D-window and participates in GPU computation at zero H2D cost.
Once $D$ entries accumulate ($D$=256, an integer multiple of the block size), the batch is evicted whole into the unpinned compute or pinned load region (realized with \texttt{cudaHostAlloc}), the side chosen by the roofline policy (\S\ref{sec:roofline}).
The write-back travels D2H---idle in our pipeline, which consumes only H2D---so full-duplex PCIe hides it inside the expert phase; hence any $D$ in 256--1024 leaves steady-state overhead unchanged,\footnote{At $D$=256, a 1000-step decode window shows flat per-token latency (TPOT, P99/P50=1.03; Qwen3, bs=8, seq=8K, A100); even at the stress bound ($D$=1024, bs=32), per-layer write-backs (37.8--134.2~MB) take 1.5--5.3~ms on Gen4 (2.9--10.3~ms on Gen3), still hidden in the expert phase.} making $D$ a routing granularity rather than a decision unit.
Placed batches never move again: exact attention keeps every entry live for the sequence's lifetime, and session-end reclamation reuses blocks in place at the same granularity.

During attention, the CPU computes over already-placed KVCache in the unpinned region, while the GPU streams the pinned region through the ping-pong region in an overwrite-reuse pattern.
The two partial outputs are merged via log-sum-exp (LSE) rescaling~\cite{flashattention3}, one join per step: the CPU returns only its partial output and normalizers---orders of magnitude smaller than raw KV---and the GPU performs the merge.
The overwrite-reuse of the ping-pong region differs from vLLM-style~\cite{vllm} repeated destroy-and-rebuild: the transfer region is created once and reused long-term---block size and skeleton never change, only the data on the blocks is replaced; the CPU side holds the full KVCache backup, so overwrites never lose data.

\begin{figure}[t]
  \centering
  \begin{subfigure}[ht]{\linewidth}
    \includegraphics[width=\linewidth]{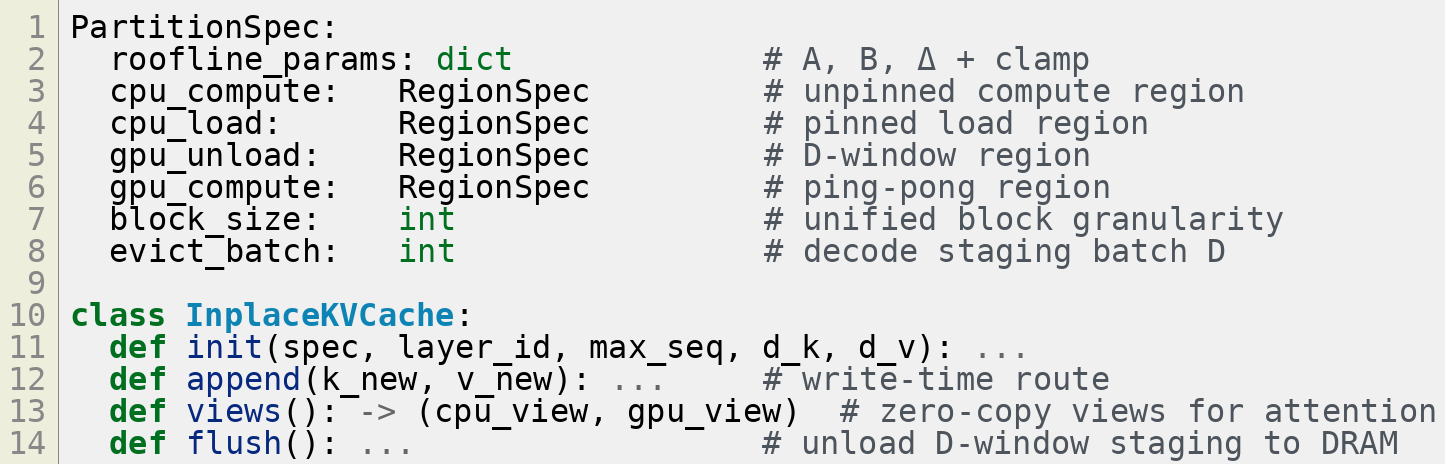}
  \end{subfigure}

\caption{InplaceKVCache interface. The \texttt{PartitionSpec} captures physical layout and roofline parameters; four thin operations enable write-time routing.}
     \label{figure5} 
\end{figure}

\subsection{Interface and Semantics}
\label{sec:interface}

To make the abstraction a reusable primitive rather than an implementation detail, we provide a framework-agnostic interface (Fig.~\ref{figure5}): a \texttt{PartitionSpec} per model/hardware (roofline parameters, the four regions, block granularity, and eviction batch size), and four operations of InplaceKVCache---\texttt{init} sizes the four regions and pre-allocates the unpinned compute, ping-pong, and D-window regions, \texttt{append} routes each entry at write time, \texttt{views} returns zero-copy views for both sides, and \texttt{flush} offloads the D-window staging to the pinned or unpinned regions in proportion during decode (\S\ref{sec:routing}).
The pinned load region instead grows batch by batch, its pinning kept off the critical path: prefill's share is pinned during model loading and warmup, and each decode batch is pinned one eviction window ahead of its use (\S\ref{sec:footprint}).

It guarantees four semantics: (i) no temporary monolithic state---\texttt{append} is the routing decision; (ii) physical isolation---the pin and layout attributes of each region are fixed by the format, while device residency follows runtime routing; each engine reads only its own region without crossing boundaries; (iii) zero-copy views---no copying or reorganization (under non-sparse attention); (iv) exactness---no data compression or sparsification beyond the model's native mechanism.
The interface is deliberately thin: four operations constitute a complete yet minimal abstraction surface.

\begin{figure}[t]
  \centering
  \begin{subfigure}[ht]{\linewidth}
    \includegraphics[width=\linewidth]{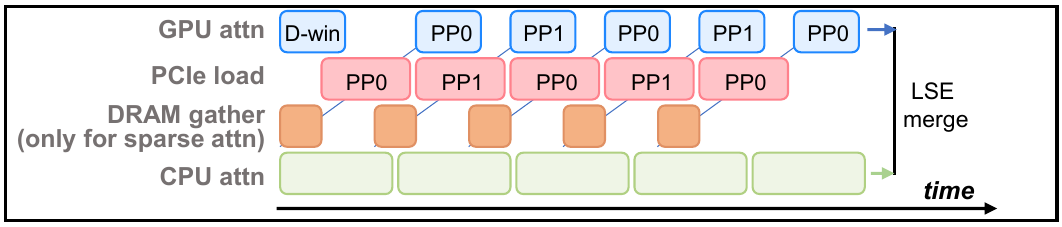}
  \end{subfigure}
\caption{Pipelined sparse-selection loading. DRAM gather is hidden within the PCIe loading, leaving no extra overhead for block-level sparsity.}
     \label{figure6} 
\end{figure}

\subsection{Sparse-Selection Compatibility}
\label{sec:sparse}

The above mechanism assumes dense full-sequence attention.
Models such as DeepSeek-V4~\cite{ds4}, however, have begun adopting sequence-dimension sparsity.
The block-aligned format supports it unchanged: any block-level top-$k$ selection yields a set of whole blocks, and a staging buffer reserved in the load region gathers them into one contiguous segment the size of the ping-pong buffer, converting the discrete transfers of sparse selection into one full-bandwidth DMA.

The gather itself is hidden by pipelining (Fig.~\ref{figure6}): measured gather bandwidth (35.1~GB/s) exceeds H2D (25.4~GB/s), so it never stalls the PCIe stream.
Sparsity granularity therefore stops at the block level: each selected block costs one \texttt{memcpy}, whereas token-level selection would drown in per-segment invocation overhead.
These mechanisms are exercised in the DeepSeek-V4 case study (\S\ref{sec:v4}).
\section{Load-Balancing Scheduling Theory}
\label{sec:scheduling}

\subsection{Sequence-Dimension Splitting}
\label{sec:splitting}

Throughout this section, let $r$ denote the CPU's share of the sequence and $L$ the already-placed sequence length.
We split the load along the sequence dimension, not the head dimension: the sequence axis makes the split ratio continuous, whereas a head-wise split is quantized by the KV-head count---coarse under GQA (four KV heads in Qwen3) and ill-defined under MLA's shared latent.
Both engines compute over all heads but on disjoint sequence ranges: the GPU handles the D-window staging segment (the latest $D$ entries, zero H2D) and the adjacent $(1-r)\cdot L$ segment streamed from the pinned region, while the CPU handles the remaining $r\cdot L$ segment in the unpinned region.
The partial outputs merge exactly via LSE rescaling; HGCA~\cite{hgca} applies the same merge across the CPU-GPU boundary but with a static split---GPU share fixed by VRAM capacity, CPU covering a sparsified remainder.
We make the split dynamic: the ping-pong region decouples the GPU share from VRAM capacity (\S\ref{sec:routing}), so $r$ is tunable over $[0,1]$ and tracks the workload (\S\ref{sec:roofline}).

\subsection{Analytic Performance Model}
\label{sec:model}

The CPU side is bandwidth-bound: attention's arithmetic intensity (0.5 MAC/byte) sits two orders of magnitude below the CPU's compute--bandwidth balance point~\cite{williams2009roofline}, so each KV token it processes incurs an effective read cost $A$.\footnote{Per KV token, attention performs $O(d)$ arithmetic on $O(d)$ bytes ($d$ the head dimension); running CPU attention with 8--48 threads shifts the measured $A$ by less than 5\% on both Qwen3 and DS2 (bs=16, seq=16K).}
Each token routed to the GPU incurs a transfer cost $B$, dominated by H2D transfer with compute hidden beneath it (both costs are per token at a given batch size).
With fixed per-step overheads $C_{\text{CPU}}$ and $C_{\text{GPU}}$ (launch and scheduling), the two engine latencies are
\begin{equation}
T_{\text{CPU}}(r) = r \cdot A \cdot L + C_{\text{CPU}}, \quad T_{\text{GPU}}(r) = (1-r) \cdot B \cdot L + C_{\text{GPU}},
\label{eq:latency}
\end{equation}
and end-to-end latency is $T = \max(T_{\text{CPU}}, T_{\text{GPU}}) + T_{\text{merge}}$; write $\Delta = C_{\text{GPU}} - C_{\text{CPU}}$.
The merge term is $r$-independent---one LSE join per step regardless of the split, measured at 0.2--1.0~ms---so it cannot affect the optimal split that Eq.~\eqref{eq:roofline} solves for.
$A$ is not a constant---it shifts with working set and cache locality---which is why the optimal share varies with workload (\S\ref{sec:load-balance-need}).

\begin{figure}[t]
  \centering
  \begin{subfigure}[ht]{\linewidth}
    \includegraphics[width=\linewidth]{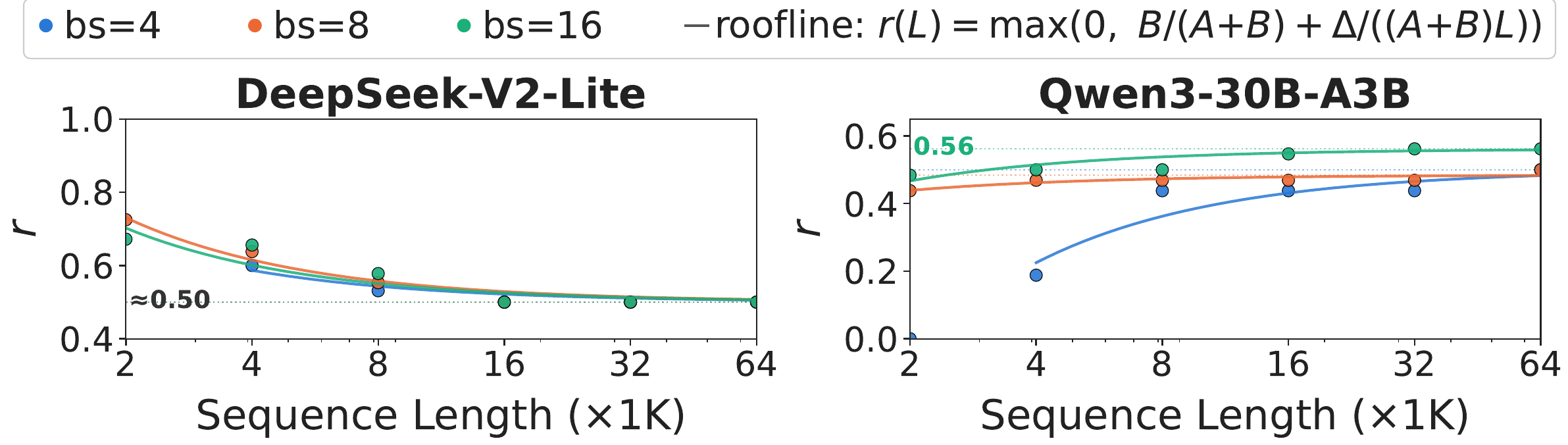}
  \end{subfigure}
\caption{The roofline optimum $r(L)$ for DeepSeek-V2-Lite and Qwen3-30B-A3B, compared with measured values.}
     \label{figure7} 
\end{figure}

\subsection{Allocating the CPU Share with Roofline}
\label{sec:roofline}

With both engines executing in parallel, load balancing requires $T_{\text{CPU}} = T_{\text{GPU}}$, yielding:
\begin{equation}
r(L) = \min\left(1, \max\left(0, \frac{B}{A+B} + \frac{\Delta}{(A+B) \cdot L}\right)\right)
\label{eq:roofline}
\end{equation}
Of the three parameters, $B$ is calibrated once, a constancy that the format itself provides: per-token volume is fixed by the model, and all GPU-bound KV moves through ping-pong blocks of fixed size, layout, and pin state (\S\ref{sec:routing}), so every transfer repeats the same shape.\footnote{Transfer size shapes achieved bandwidth: on Gen4, 192~KB transfers reach 8.7~GB/s, 1~MB 22.2~GB/s, and 8~MB the 25.4~GB/s used in calibration; the ping-pong buffers (37.8--134.2~MB, Table~\ref{tab:footprint}) sit firmly on this plateau.}
$A$, by contrast, evolves with the workload and must be tracked online.
In the long-context regime $B/(A+B)$ determines the converged value, while $\Delta/((A+B) \cdot L)$ characterizes the transition at short sequences; clamping to 0 or 1 handles extreme configurations that require no load balancing (\S\ref{sec:eval-portability}).
Fig.~\ref{figure7} shows $r(L)$ for DS2 and Qwen3: DS2 has positive $\Delta$ and $A$ varies little with bs, so $r(L)$ decays to the converged value of 0.50; Qwen3 has negative $\Delta$ and $A$ degrades with bs, so $r(L)$ transitions gradually and the converged value rises with bs.

Operationally, $r(L)$ guides D-window eviction during decode: rather than freezing a fixed share, we hold the marginal share constant---successive eviction batches are routed so that the running fraction of CPU-placed batches matches $p = B/(A+B)$, the converged value of $r(L)$.
During D-window staging, the CPU-side component latency per decode step is sampled via a sliding average and $A$ is back-computed; 256 steps across multiple layers yield an accurate estimate.
The initial share $R(L_0) = r(L_0)$ at admission is given by offline calibration---scanning roofline parameters at different data volume scales for the target model, which takes only minutes---so adaptation starts on target and the online loop only tracks drift.
Consequently, convergence is a matter of batch geometry, not
statistics: the cumulative share
$R(L) = p + (R(L_0) - p)\cdot L_0/L$ moves toward the marginal
share $p$ by $D/L$ per batch, so an initial share error decays
within the first few batches of decode---a vanishing fraction of
a long decode.\footnote{The deployed share stays within 3\% of the empirically measured optimum for long sequences and within 5.1\% (block quantization) for short ones; calibration, online feedback, and cross-platform validation are in \S\ref{sec:eval-portability}.}
Thereafter, each D-window updates $A$ with the latest sliding average and adjusts the share accordingly; in the quasi-static regime where $A$, $B$, and $\Delta$ are approximately constant, $R(L)$ tracks $r(L)$ batch by batch, adapting through new placements alone.
\section{Implementation}
\label{sec:implementation}

\subsection{CPU Kernel for In-Place Attention}
\label{sec:cpu-kernel}

CPU-side attention is executed by a custom fused AVX-512 kernel.
Decode attention is memory-bound, and two designs in the kernel carry the format's guarantees into the compute path: zero-copy consumption of partitioned views, and compressed-MLA reuse.

\textbf{Zero-copy consumption of partitioned views.}
The kernel directly indexes CPU-side K/V slices via stride-aware addressing, never requiring contiguous copies; scores are never materialized---each head-group's score vector lives only in registers, with online softmax~\cite{flashattention3} maintaining the running max and normalization term, and fp32 output accumulated in L1 cache.

\textbf{Compressed-MLA variant.}
In DeepSeek-V2~\cite{deepseekv2}, KV is cached as a compressed latent rather than per-head K/V, so attention cannot consume it directly.
The kernel exploits MLA's decoupled structure, reducing computation to a q-side projection and an output-side expansion. Multiple heads within a segment (an 8-token tile) share one compressed latent, so repeated accesses in both the score and output phases hit L1 and each latent is read from DRAM only once.

\subsection{Framework Integration}
\label{sec:framework}

WriteScope runs on a hybrid runtime that reuses Hugging-

\noindent Face Transformers~\cite{huggingface} as the model skeleton---module wiring and the sampling loop---and replaces every compute path: custom attention kernels (\S\ref{sec:cpu-kernel}), expert execution, and KVCache management.
On the expert side, a CPU--GPU hierarchical scheduler performs dynamic hot/cold ranking~\cite{layerscope,hybridmoe} on the IPEX~\cite{ipex} AVX-512 backend, and KVCache and expert-weight budgets are governed by a single joint allocation model.
The system core comprises approximately 2.2K lines of Python (InplaceKVCache routing and the roofline scheduler), a 1.2K-line C++ backend (CPU attention, AVX-512), and a 2.1K-line DeepSeek-V4 sparse kernel; all of it builds on a shared 5K-line substrate---expert hot/cold scheduling and per-model adaptation---that also hosts the four reproduced baselines (Appendix~\ref{app:baseline}).
Integration into the skeleton requires only a narrow interface (\S\ref{sec:interface}): routing hooks replace the attention module's KVCache \texttt{append} calls, and the \texttt{PartitionSpec} is generated offline by the roofline model before prefill, then updated continuously via online feedback during decode.

Realizing WriteScope as an incremental extension of vLLM or SGLang runs into a structural mismatch: both fix expert and KVCache budgets statically at initialization (\texttt{gpu\_memory}

\noindent \texttt{\_utilization}) and offload weights only at layer granularity, so the statically reserved KVCache pool caps the effective context length (\S\ref{sec:spill-object}), whereas WriteScope relies on a joint allocation model that adjusts the expert-resident fraction against the KVCache budget at runtime---which neither framework's extension points natively support.
We estimate such a port at roughly 5K lines of code, mostly expert-side machinery rather than the KV format.
\section{Evaluation}
\label{sec:evaluation}

\subsection{Experimental Setup}
\label{sec:eval-setup}

\textbf{Hardware.}
The primary platform is an NVIDIA A100 PCIe GPU (PCIe Gen4 $\times$16, VRAM constrained to 32~GB), with a single Intel Xeon Gold 5318Y (24 cores) and 502~GB of host memory.
The portability validation platform (\S\ref{sec:eval-portability}) is an NVIDIA V100-32GB (PCIe Gen3 $\times$16) on the same host CPU.

\textbf{Models.}
We evaluate three MoE models (DeepSeek-V2-Lite~\cite{deepseekv2} 27-layer MLA, Qwen3-30B-A3B~\cite{Qwen3} 48-layer GQA, Mixtral-8$\times$7B~\cite{mixtral} 32-layer GQA, $d_h$=128, bf16) and DeepSeek-V4-Flash-Int8~\cite{ds4}, which mixes Compressed Sparse Attention (CSA) and Heavily Compressed Attention (HCA) (43-layer, $d_h$=512, weights INT8, KV bf16), for a sparsity-generalization case study (\S\ref{sec:v4}).

\textbf{Baselines.}
Two tiers: three inference frameworks run natively (KTransformers~\cite{ktransformers} 0.4.1, SGLang~\cite{sglang} 0.5.4, vLLM~\cite{vllm} 0.14) and four attention baselines chosen to span the execution-placement spectrum: GPU-centric (SolidAttention~\cite{SolidAttention}: DRAM storage, PCIe streaming load to GPU, sparsified), CPU-centric with a GPU-resident share (ScoutAttention~\cite{scoutattention}: GPU-resident top-$k$ blocks, CPU layer-ahead pre-computation, sparsified), staged CPU-logits/GPU-aggregation (HybridGen~\cite{hybridgen}), and fully CPU-resident (FastDecode~\cite{fastdecode}).
The four attention baselines target dense models and have no MoE expert execution path.
Each baseline is therefore reproduced on our runtime under the fairness protocol of Appendix~\ref{app:baseline} (deviations labeled by direction; validated against original reported numbers, Table~\ref{tab:deviations}), and all baselines and WriteScope share one runtime, one expert scheduler, and one measurement harness---the runtime is a controlled constant.
WriteScope computes exact full attention throughout, while the two sparse baselines retain 80\% of KV, so the reported speedups are achieved while processing strictly more data.

\textbf{Workload.}
Prefill inputs are drawn from real conversations in ShareGPT~\cite{sharegptv3}.
Measurement uses a 2K-token prompt followed by decode generation to the target length; the 8K/32K points report cumulative end-to-end throughput, while the 2K point reports steady-state performance over 256 output tokens after reaching the target.
Each configuration is run independently 10 times reporting the mean; 95\% confidence intervals for speedup (bootstrap) are within $\pm$5\%.

\textbf{KV resident budget.}
WriteScope and the four reproduced baselines are capped at a 2~GB VRAM-resident KVCache budget (0.3~GB for DS2 thanks to MLA compression); the three inference frameworks, whose offloading capability is limited, are granted 4~GB.\footnote{Under the same 2~GB budget, the frameworks would yield only a few bs=4 data points---too sparse to analyze.}
The comparison is thus conservative for our system: the frameworks run with twice WriteScope's budget.

\begin{figure}[t]
  \centering
  \begin{subfigure}[ht]{\linewidth}
    \includegraphics[width=\linewidth]{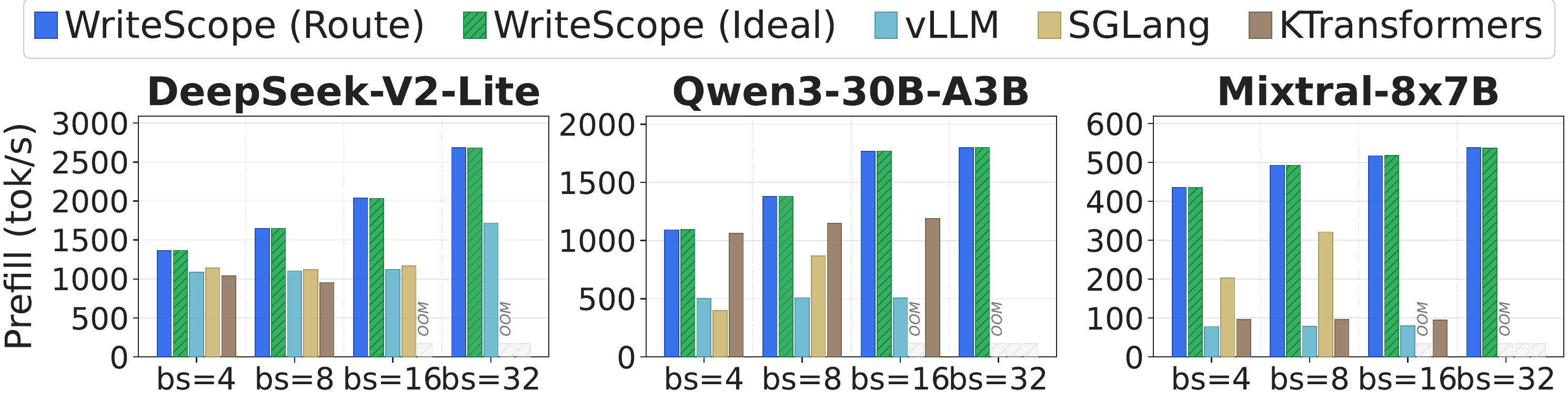}
  \end{subfigure}
  \caption{Prefill throughput (2K prompts). WriteScope (Route vs.\ Ideal) shows write-time routing is free; mainstream frameworks OOM at scale.}
     \label{figure8} 
\end{figure}

\begin{figure*}[t]
  \centering
  \begin{subfigure}[ht]{\linewidth}
    \includegraphics[width=\linewidth]{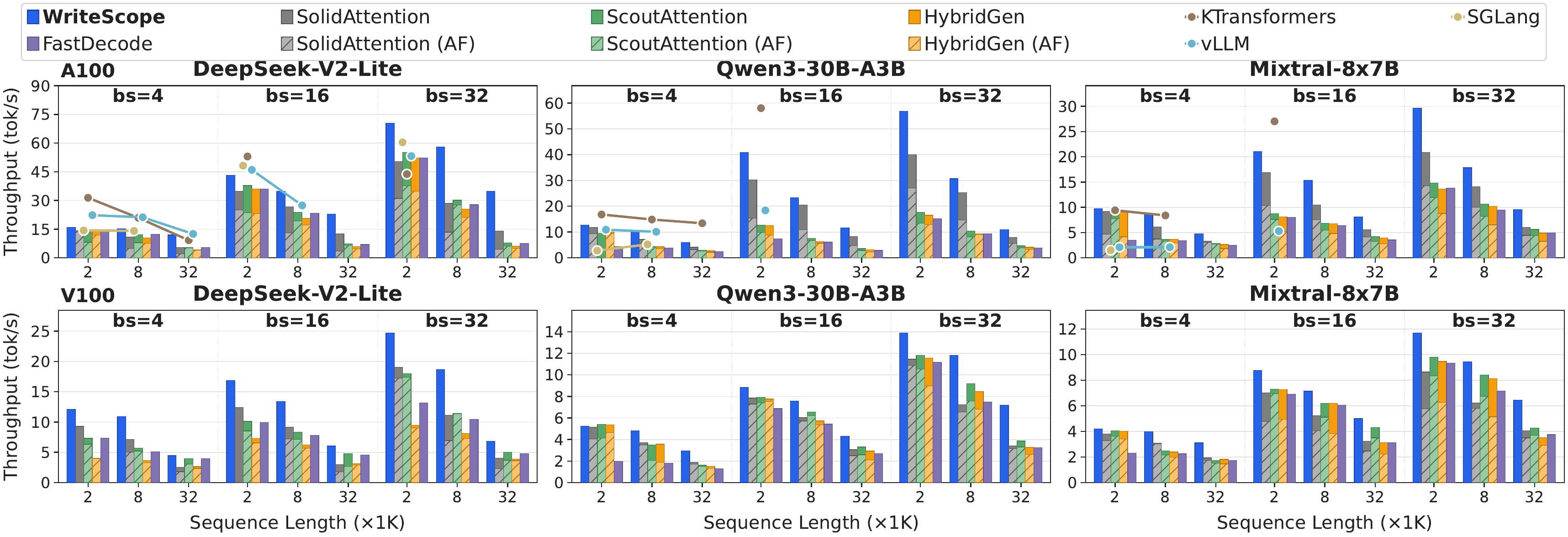}
  \end{subfigure}
\caption{End-to-end decode throughput. For each AF-overlap baseline, the solid and hatched bars at the same x-coordinate are its non-AF and AF variants. Mainstream frameworks are shown as points/lines (missing = OOM).}
     \label{figure9} 
\end{figure*}

\subsection{Prefill: Throughput and Capacity Robustness}
\label{sec:eval-prefill}

Write-time routing is part of prefill, and its cost must be measured.
Fig.~\ref{figure8} reports steady-state prefill throughput at 2K prompts for the three models at batch sizes (bs) 4, 8, 16, and 32, comparing five engines: WriteScope (Route, full write-time routing), WriteScope (Ideal, control without write-time routing), vLLM, SGLang, and KTransformers.

Three observations.
First, write-time routing is free: Route and Ideal are nearly identical across all configurations, e.g., 1365 vs.\ 1366 tok/s for DS2 at bs=4, with difference $\leq$1 tok/s---direct system-level evidence of the write-time principle (\S\ref{sec:timing}).
Second, capacity robustness: the three frameworks keep at-

\noindent tention KV fully GPU-resident, and once total demand exceeds their spill tolerance they OOM---at bs$\geq$16 in Fig.~\ref{figure8}; WriteScope completes all configurations, and the capability boundary itself is part of the contribution.
Third, in the spill regime WriteScope leads on every runnable configuration, and its throughput keeps rising with bs---DS2 grows from 1365 tok/s at bs=4 to 2684 at bs=32 (2$\times$ at 8$\times$ batch)---whereas budget-constrained frameworks plateau earlier; the gap is largest on strictly-spilled Mixtral (WriteScope 435--538 vs.\ vLLM 78--80 tok/s, a 5.4--6.9$\times$ gap).

WriteScope supports long-prompt prefill through layer-by-layer KVCache eviction and flexible expert offloading: at bs=4 and seq=32K, TTFT (time to first token) for DS2, Qwen3, and Mixtral is 51, 71, and 132~s, corresponding to 2570, 1850, and 990 tok/s.

\subsection{End-to-End Decode Throughput}
\label{sec:eval-decode}

Fig.~\ref{figure9} gives end-to-end decode throughput on both platforms and all three models.
We report speedup as WriteScope's throughput divided by the baseline's.

First, in the seq$\geq$8K regime (across three models, batch sizes 4, 16, and 32), WriteScope's geometric-mean speedup over each baseline is: on A100, 1.52$\times$ over SolidAttention, 2.22$\times$ over ScoutAttention, 2.48$\times$ over HybridGen, 2.51$\times$ over FastDecode; on V100, 1.55$\times$, 1.43$\times$, 1.67$\times$, 1.71$\times$ respectively.
The platform shift is itself informative: with the same CPU, V100's slower PCIe and GPU narrow our margin over the CPU-centric baselines, yet the margin over GPU-centric SolidAttention holds (1.52$\rightarrow$1.55$\times$). The halved PCIe should widen the attention-side gap, but V100's slower GPU also drags expert execution on both sides, compressing the end-to-end difference; WriteScope simply shifts its balance point toward the CPU (\S\ref{sec:eval-portability}).

Second, WriteScope's advantage amplifies with sequence length.
On A100, with DS2 and bs=32, the speedup over Sol-

\noindent idAttention grows from 1.40$\times$ at 2K to 2.49$\times$ at 32K, and over ScoutAttention from 1.28$\times$ at 2K to 4.42$\times$ at 32K.
As load grows, single-side bottlenecks deepen while load balancing keeps both engines busy---the larger the load, the more valuable write-time partitioning becomes.

Third, no single fixed strategy consistently outperforms. SolidAttention's pure PCIe-loading approach suffers most on V100, where halved PCIe bandwidth magnifies its per-step H2D cost; FastDecode's fully CPU-resident strategy falls short on short sequences, where GPU-resident KV could have handled a substantial fraction of the attention computation at no transfer cost; ScoutAttention and HybridGen sit in between at different intermediate points—none covers the full spectrum of model architectures and sequence lengths.

Fourth, AF overlap is universally worse than the non-overlap variant, and the gap widens with sequence length as expert contention grows.
ScoutAttention exploits AF overlap to outperform HGCA on dense models, but under MoE workloads the reverse holds: the de-overlapped, HGCA-equivalent variant wins (Appendix~\ref{app:baseline}).

Finally, taking mainstream frameworks as reference: with a higher KVCache VRAM budget, vLLM, SGLang, and KTransformers even exceed WriteScope throughput at some small-load points, confirming that GPU-resident KV is efficient for short context but fails broadly at long context.

\subsection{Ablation Study}
\label{sec:eval-ablation}

\begin{figure}[t]
  \centering
  \begin{subfigure}[ht]{\linewidth}
    \includegraphics[width=\linewidth]{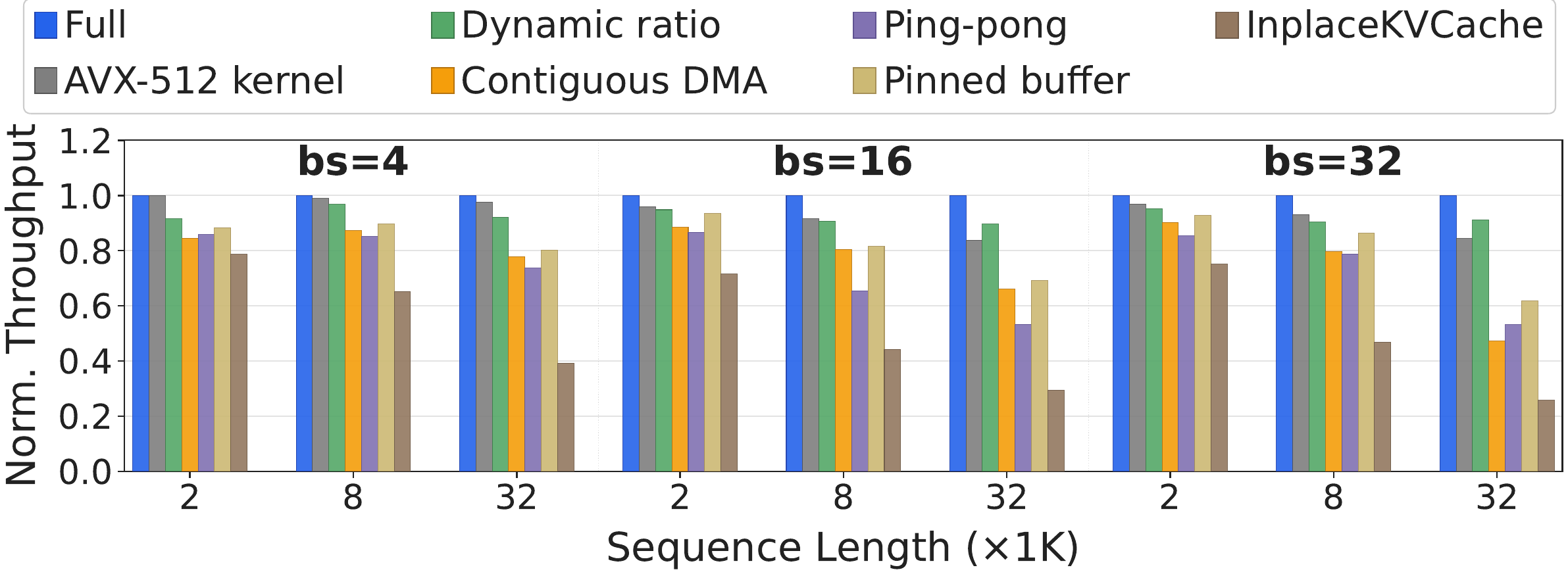}
  \end{subfigure}
\caption{Ablation on DeepSeek-V2-Lite, normalized to the full system. Every other series replaces the design point it names with the baseline implementation; lower is worse.}
     \label{figure10} 
\end{figure}

Fig.~\ref{figure10} ablates one design point at a time on DS2, replacing it with the corresponding baseline implementation and reporting normalized end-to-end throughput.

\textbf{Replace custom AVX-512 kernel with IPEX oneDNN fused GEMM~\cite{ipex,onednn}:}
0.84--1.00 normalized throughput, the mildest arm, since the replacement is itself an industrial-grade optimized library, showing that kernel engineering's end-to-end contribution is bounded.

\textbf{Replace the dynamic ratio with the fixed converged $r$=0.50:}
0.89--0.96 normalized throughput---the optimal share drifts with sequence length (Fig.~\ref{figure7}), a frozen share matches it at only one point, while the marginal-share policy tracks it throughout.
This arm is not the static partitioning of prior work~\cite{scoutattention,hgca}, whose GPU-resident fraction is bounded by VRAM capacity (DS2's 2~GB budget is only 6\% of the total KV at seq=32K, bs=32) for lack of an overwrite-reuse mechanism like our ping-pong region; it isolates the share alone, showing that roofline-guided dynamic allocation is necessary on top of InplaceKVCache.

\textbf{Replace contiguous DMA with strided DMA:}
0.47--0.90 normalized throughput.
Degradation amplifies with sequence length and bs, quantifying the stride collapse of \S\ref{sec:strided}.

\textbf{Replace ping-pong overwrite-reuse with per-step rebuild:}
0.53--0.87 normalized throughput.
This quantifies the per-step allocation, deallocation, and synchronization cost that overwrite-reuse avoids.

\textbf{Replace pinned buffer with unpin:}
0.62--0.94 normalized throughput.
This quantifies the seesaw of \S\ref{sec:pinning}: unpinning the H2D source forfeits the transfer half of the trade.

\textbf{Replace InplaceKVCache with DynamicCache:}
0.26--0.79 normalized throughput, the heaviest degradation---it re-walks the full prepare pipeline of allocation, splicing, pin, chunked H2D every step, lacking the bundling effect of write-time routing. It is the combined projection of the three measurements in \S\ref{sec:why-monolithic} (pin seesaw, stride collapse, non-amortizable timing)---the very path a runtime wrapper over a monolithic pool must walk (\S\ref{sec:format-conclusion}).
Its degradation exceeds any single-factor arm, showing that each design point of the four-region format is not decorative.

The pin seesaw, stride collapse, and reallocation overhead are hardware-layer phenomena, hence model-agnostic; only the AVX-512-kernel and Dynamic-ratio arms show model dependence. On Qwen3, the Dynamic-ratio arm degrades similarly (0.77--0.95), and the AVX-512-kernel arm remains bounded (0.84--1).

\subsection{Composing with DeepSeek-V4's Sparsity}
\label{sec:v4}

DeepSeek-V4 mixes two attention mechanisms with sharply different KV compression ratios.
At bs=32 $\times$ seq=64K, for example, the GPU permanently holds only 0.5~GB of compact state---window KV (4.2~MB $\times$ 43 layers, 180~MB) and the HCA layers' 128$\times$-compressed KV (16.8~MB $\times$ 20 layers, 336~MB)---whereas the 21 CSA layers, at only 4$\times$ compression, each leave 671.1~MB in CPU memory (536.9~MB compressed KV + 134.2~MB block-selection indexer cache), 14.1~GB in total.
KVCache management therefore reduces to the CSA layers.
This case study validates the composition of write-time routing with model-native sparsity: all measurements respect the model's CSA/HCA configuration and apply no external resparsification.

\begin{figure}[t]
  \centering
  \begin{subfigure}[ht]{\linewidth}
    \includegraphics[width=\linewidth]{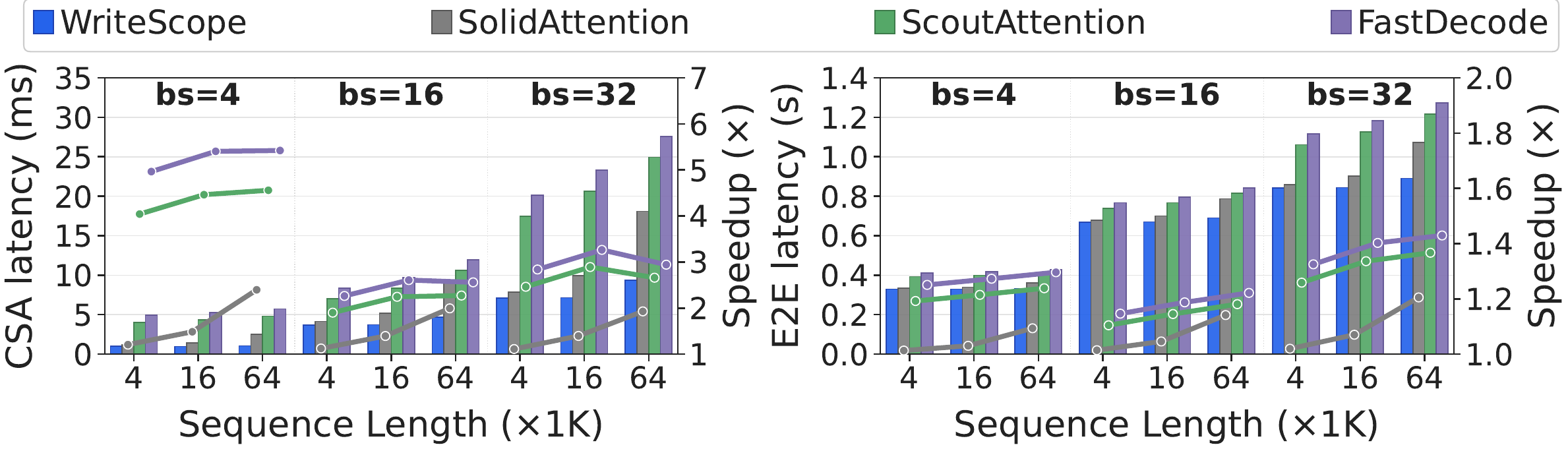}
  \end{subfigure}
\caption{Single-layer CSA latency (left) and end-to-end latency (right) in DeepSeek-V4. Lines report WriteScope's speedup over each baseline on the right axis.}
     \label{figure11}
\end{figure}

Fig.~\ref{figure11} compares single-layer CSA attention across four strategies (WriteScope, SolidAttention, ScoutAttention, FastDecode): WriteScope achieves the lowest latency at every configuration, with advantages of 1.11--2.40$\times$ over SolidAttention, 1.90--4.56$\times$ over ScoutAttention, and 2.26--5.42$\times$ over FastDecode, growing with sequence length.

At the end-to-end level, the four strategies run on the same framework sharing identical expert and HCA+dense execution paths, differing only in the CSA attention mechanism---so end-to-end differences are attributable to the CSA segment.
In the seq$\geq$16K regime, WriteScope's geometric-mean speedups over SolidAttention, ScoutAttention, and FastDecode are 1.10$\times$, 1.24$\times$, and 1.30$\times$; at the heaviest bs=32 $\times$ seq=64K configuration, they rise to 1.21$\times$, 1.37$\times$, and 1.43$\times$.

HybridGen is absent from this comparison: CSA compresses K and V into a single shared representation, leaving no K/V boundary for its CPU-GPU split.
DeepSeek-V4 marks the boundary of composability: sparsity shrinks the compute volume while changing neither physical ownership nor compute engine---the very properties write-time routing relies on.
WriteScope exploits this independence: exact attention by default, composition with model-native sparsity, and external sparsification as an orthogonal user option (any block-level selector plugs into block-aligned regions without format changes, \S\ref{sec:sparse}).

\subsection{Portability and Applicability Boundary}
\label{sec:eval-portability}

\begin{figure}[t]
  \centering
  \begin{subfigure}[ht]{\linewidth}
    \includegraphics[width=\linewidth]{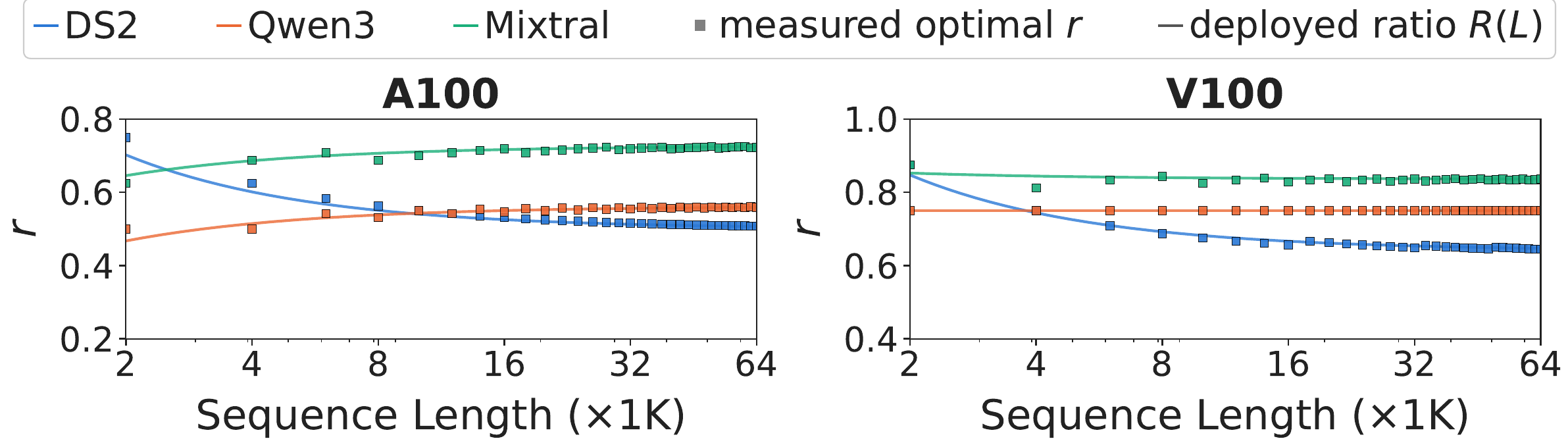}
  \end{subfigure}
  \caption{Deployed share $R(L)$ (curves, roofline-guided) vs.\ the measured optimal $r$ (points: per 2K of context, swept over the following 256 decode steps), on A100 and V100 (bs=16).}
     \label{figure12}
\end{figure}

To verify that the roofline model of \S\ref{sec:roofline} generalizes across hardware and models, we compare the share the system actually deploys---the roofline-guided trajectory---against the empirically optimal balance point measured at each data volume, on both A100 and V100 across three models (Fig.~\ref{figure12}, bs=16).
The two stay close: at the start of decode the deviation is 0.3\%--4.5\%; for long sequences ($\geq$8K) it converges to $\leq$3\%; short sequences (2K--4K) add a block-quantization deviation of at most 5.1\%---one $D$=256 eviction batch occupies 12.5\% of a 2K context---that attenuates as $L$ grows.
The model-specific shapes of the curves are reproduced on both platforms, with V100's converged values uniformly higher because lower BW$_{\text{H2D}}$ makes the GPU side more expensive---the model's guidance lands on the measured optimum wherever the platform moves it.

\begin{figure}[t]
  \centering
  \begin{subfigure}[ht]{\linewidth}
    \includegraphics[width=\linewidth]{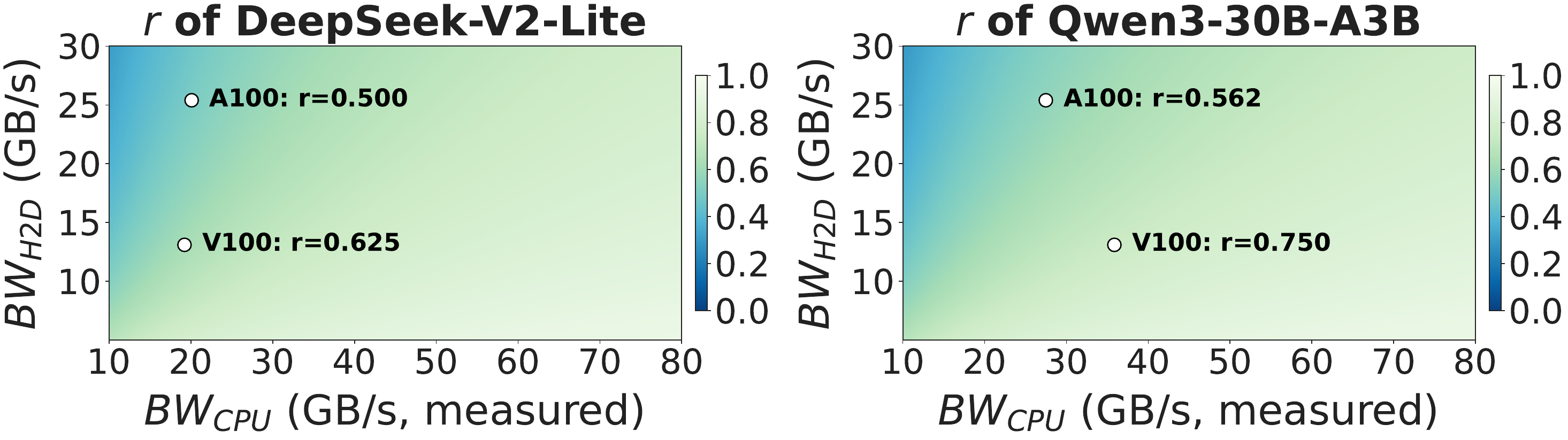}
  \end{subfigure}
  \caption{Extrapolated converged $r$ across CPU bandwidth and PCIe BW$_{\text{H2D}}$; A100 and V100 anchor the surface, and regions far from both anchors are indicative only.}
     \label{figure13}
\end{figure}

With both anchors validated, Fig.~\ref{figure13} extrapolates the converged $r$ across the hardware plane, with measured CPU bandwidth and PCIe BW$_{\text{H2D}}$ as axes and A100/V100 as the anchor points.
The surface delineates our applicability boundary: where the two bandwidths are vastly imbalanced, $r$ clamps to 0 or 1 and the cache degenerates to a pure GPU- or CPU-centric regime; where they are comparable and VRAM forces spilling---exactly the single-box setting---$r$ falls strictly between 0 and 1 and dynamic balancing pays.

\subsection{System Resource Footprint}
\label{sec:footprint}

Table~\ref{tab:footprint} gives the memory footprint of WriteScope's four regions (measured at seq=32K, bs=32).
GPU-side residency comprises only two regions---the ping-pong region and the D-window region---totaling 0.29/0.85/1.18~GB at the deployed $D$=256, less than 1\% of the full KV at the same configuration.
The CPU side is split by write-time placement into unpinned compute and pinned H2D source, whose sum equals the full KV.
The design thus decouples GPU VRAM from sequence length: for a given model, GPU residency grows with $D \times \mathrm{bs}$---and $D$ is a knob should the budget tighten---leaving CPU memory as the sole capacity bound for long context.

Pinning is routed in time as well: each D-window batch's \texttt{cudaHostAlloc} is issued when its predecessor lands, overlapping the page-lock work with the 256-step eviction window.
At 0.66~ms/MB, even the largest single allocation (711~ms, 1074~MB) is covered by this lookahead by over two orders of magnitude (256 steps at bs=32 take $\geq$100~s).
Unlike the per-step rebuild of \S\ref{sec:timing}, each block is pinned once in its lifetime, ahead of consumption, and over new bytes only---the write-time principle applied to its own metadata.

\begin{table}[t]
\centering
\caption{WriteScope memory footprint (seq=32K, bs=32).}
\label{tab:footprint}
\resizebox{\linewidth}{!}{
\begin{tabular}{lcccc}
\toprule
\textbf{Model} & \textbf{Ping-pong} & \textbf{D-window} & \textbf{Unpinned} & \textbf{Pinned} \\
\midrule
DS2 & 37.8~MB & 255~MB & 15.95~GB & 16.66~GB \\
Qwen3 & 67.1~MB & 805~MB & 61.95~GB & 41.13~GB \\
Mixtral & 134.2~MB & 1074~MB & 103.08~GB & 34.36~GB \\
\bottomrule
\end{tabular}
}
\end{table}

\subsection{Multi-User Concurrency: Interference Analysis}
\label{sec:concurrency}

The preceding evaluation assumes a single user issuing multi-request long-context workloads; this section supplements preliminary measurements under multi-user concurrency.
Two users start their multi-request tasks with staggered launches and overlapping execution, sharing the same single-box single-GPU platform.
Each user's latency under exclusive resources serves as the reference (solo), reported alongside the steady-state share $r$ that roofline self-adaptation converges to in that setting.
On this basis we compare two strategies in Table~\ref{tab:interference}: (1) \emph{fixed $r$}, freezing the solo $r$; (2) \emph{runtime adaptive}, independently monitoring $A$ and $B$---contention now perturbs $B$ as well---and adjusting $r$.

\begin{table}[t]
  \centering
  \caption{Concurrent interference and adaptive mitigation on Qwen3-30B-A3B (two concurrent pairs, ms/step).}
  \label{tab:interference}
  \resizebox{\linewidth}{!}{
  \begin{tabular}{lccccc}
  \toprule
  \textbf{Request} & \textbf{solo $r$} & \textbf{adaptive $r$} & \textbf{solo } & \textbf{fixed-$r$ } & \textbf{adaptive } \\
  \midrule
  bs8/8K   & 0.469 & 0.375--0.438  & 499  & 779  & 651 \\
  bs8/16K  & 0.469 & 0.18--0.469   & 664  & 959  & 827 \\
  \hline
  bs4/8K   & 0.438 & 0--0.28       & 386  & 1010 & 436 \\
  bs16/16K & 0.547 & 0.18--0.547   & 1095 & 1253 & 1134 \\
  \bottomrule
  \end{tabular}
  }
  \end{table}

Under contention, \emph{fixed $r$} degrades severely---it cannot perceive congestion on the shared CPU and PCIe---with bs4/8K latency inflating 2.6$\times$ over solo.
\emph{Runtime adaptation} reads the inflated costs and retunes the roofline target: the deployed share follows the congestion---collapsing from 0.438 to 0--0.28 at bs4/8K---and latency falls accordingly, by 57\% (bs4/8K, 1010$\rightarrow$436~ms) and 16\% (bs8/8K, 779$\rightarrow$651~ms) over fixed-$r$.
In the tightest configuration (bs16/16K), adaptive latency sits only 4\% above solo versus 14.4\% under fixed-$r$, and $r$ recovers rapidly once the competing request completes---the batch-geometry convergence of \S\ref{sec:roofline} under the harshest drift the system faces.

The mechanism is agnostic to the contention source---$A$ and $B$ capture any aggregate cost change, whether from concurrent experts, attention, or non-inference tasks.
It is the same loop as in \S\ref{sec:roofline}---measure the current costs, adjust $r$---except that the variation now comes from concurrent competition rather than sequence growth.
Viewed on the surface of Fig.~\ref{figure13}, contention shifts the platform's effective operating point: both bandwidths drop, the CPU's more so in our runs, which is why the adapted $r$ falls below solo---adaptation follows the surface to wherever the point lands.
\section{Related Work}
\label{sec:related}

\textbf{KVCache abstractions} have evolved along a single dimension, capacity.
DynamicCache and StaticCache~\cite{huggingface} manage growth and reallocation; PagedAttention~\cite{vllm} and vAttention~\cite{prabhu2025vattention} manage fragmentation and virtual-physical mapping; RadixAttention~\cite{sglang} shares prefixes across requests (extended by HiCache to a three-tier hierarchy).
KVDrive~\cite{kvdrive}, Tutti~\cite{tutti}, eLLM~\cite{ellm}, and Bidaw~\cite{hu2026bidaw} optimize where generated KV is stored; Mooncake~\cite{mooncake}, AttentionStore~\cite{attentionstore}, and CacheGen~\cite{cachegen} organize KV tiers across devices, host memory, and the network; transport layers such as LMCache/NIXL~\cite{lmcache} pipeline tier-to-tier movement, but every retrieved byte returns to the GPU for attention.
All decide where and how much KV is stored, never what physical state bytes occupy or which engine consumes them.

\textbf{Attention offloading} falls into two camps.
GPU-centric systems (FlexGen~\cite{flexgen}, ZeRO-Inference~\cite{zeroinference}, HeadInfer~\cite{luo2025headinfer}, SolidAttention~\cite{SolidAttention}) keep attention on the GPU and stream KV over PCIe, cutting volume by sparsity; CPU-centric systems (NEO~\cite{neo}, ScoutAttention~\cite{scoutattention}, HybridGen~\cite{hybridgen}, Fluxion~\cite{yao2026fluxion}) make the CPU the primary engine and offload a VRAM-bounded share.
HGCA~\cite{hgca} splits attention across both devices with LSE merge, but its split is static and capacity-dictated (\S\ref{sec:splitting}).
Our distinction is the decision layer: content-level selection, however often re-selected, never tunes the engine split against measured costs; WriteScope fixes residency in the format at write time and drives the split with an online roofline policy.

\textbf{MoE inference} optimizes the expert side: precision and prefetching (MoE-APEX~\cite{MoE-APEX}, FineMoE~\cite{finemoe}, FloE~\cite{floe}), CPU-GPU expert orchestration (Fiddler~\cite{fiddler}, KTransformers~\cite{ktransformers}, LayerScope~\cite{layerscope}), disaggregation (JANUS~\cite{janus}), and static or content-level placement (PowerInfer~\cite{PowerInfer,powerinfer2}, InfiniGen~\cite{InfiniGen}, llama.cpp~\cite{llamacpp}).
In all of them attention stays on the GPU, and their expert traffic saturates the CPU and PCIe during FFN---the two lines do not compose with AF-overlap attention offloading (\S\ref{sec:af-break}).
WriteScope is the first to bring attention's physical placement into the MoE design space.

\section{Discussion and Limitations}
\label{sec:discussion}

\textbf{\textit{Applicability boundary.}}
Our mechanism provides the greatest value in single-box, memory-constrained MoE long-context inference, only then does physical partitioning (the write-time principle) graduate from implementation detail to first-class semantics.
If VRAM is abundant, or if CPU and PCIe bandwidth are vastly imbalanced, the KVCache degenerates to an extreme GPU- or CPU-centric regime; the roofline surface of \S\ref{sec:eval-portability} delineates this boundary quantitatively.
KVCache quantization is likewise an orthogonal, user-level choice: it only shrinks the per-token byte count---regions, routing, and block granularity are unchanged, and the roofline recalibrates $B$ (\S\ref{sec:roofline}).

\textbf{\textit{Why has this been overlooked?}}
The absence of physical partitioning from mainstream KVCache abstractions is not accidental: in multi-GPU scenarios, VRAM is abundant; in the dense-model era, the need for load balancing was masked by AF overlap and sparsification.
This absence has also given rise to behaviors now considered canonical: GPU-resident recent KV with sparsification to reduce I/O, and offloading designed as reactive pressure relief rather than proactive management.
This paper demonstrates that GPU VRAM can be overwrite-reused within a single attention phase; what makes each offload--load cycle affordable is not the scheduling but the format beneath it---pinned sources, preserved contiguity, fixed transfer shapes.

\textbf{\textit{Limitations and future work.}}
Our target scenario is exclusive single-user access to a single-GPU system; multi-user concurrency (\S\ref{sec:concurrency}) is a feasibility validation, not a recommended deployment mode.
On hardware coverage, consumer Gen4 cards (RTX 3090/4090) share A100's PCIe generation; their converged $r$ therefore lies near the A100 anchor on the roofline surface of Fig.~\ref{figure13}. Gen5 platforms (RTX 5090) offer higher PCIe bandwidth, shifting the converged $r$ toward smaller CPU shares.
In the future, we plan to extend WriteScope to jointly schedule expert prefetching and attention placement under a unified roofline policy, managing their shared contention for CPU and PCIe bandwidth.

\section{Conclusion}
\label{sec:conclusion}

Single-GPU long-context MoE inference spills the KVCache to CPU, making dynamic load balancing between CPU-side computation and GPU-side loading on the same cache the core requirement.
We trace its long-standing absence to the data-structure level of existing KVCache abstractions and propose InplaceKVCache, the first KVCache format that fixes each byte's physical residency at write time; paired with a roofline-driven online policy, load balancing reduces to pure scheduling.
On three MoE models, WriteScope achieves geometric-mean decode speedups of 1.5$\times$--2.5$\times$ (A100) and 1.4$\times$--1.7$\times$ (V100) over four reproduced baselines in the $\geq$8K regime, sustains end-to-end inference at the 1M-token aggregate scale where mainstream frameworks fail, and composes with model-native sparsity (DeepSeek-V4)---moving long-context MoE inference from a multi-GPU-server capability to a practical configuration on single-GPU systems.

\bibliographystyle{ACM-Reference-Format}

\bibliography{bib2}
\appendix
\section{Baseline Mechanism Alignment}
\label{app:baseline}

All reproduced baselines run on the WriteScope runtime, sharing the same MoE expert scheduling backend, CPU-GPU execution pipeline, and measurement harness; the runtime is a controlled constant, so end-to-end differences isolate the KV-management mechanisms (\S\ref{sec:eval-setup}).
Reproduction follows four rules:
(R1)~the original core mechanism is preserved verbatim; only what the workload structurally requires is changed (storage medium, platform-specific memory tiers, node count);
(R2)~every deviation from the original is listed in Table~\ref{tab:deviations} with the direction it favors;
(R3)~each reproduction is validated against the original paper's reported numbers on a reproducible configuration, using hardware-normalized anchors (Table~\ref{tab:repro-validation});
(R4)~no baseline number in this paper comes from a native implementation that cannot execute the workload.

\begin{table*}[t]
\centering
\caption{Deviations from the original baselines and the direction each favors.}
\label{tab:deviations}
\footnotesize
\begin{tabular}{@{}lllp{0.52\textwidth}@{}}
\toprule
Baseline & Deviation & Favors & Rationale \\
\midrule
SolidAttention & SSD$\rightarrow$DRAM & baseline & removes the I/O bottleneck; PCIe streaming hits the H2D peak (25.4~GB/s) \\
Solid./Scout. & KV retention set to 80\% & mixed & the quality-retention point reported in the original papers; keeps the comparison on transfer/compute mechanisms \\
HybridGen & CXL tier unavailable & neutral & 502~GB host DRAM already holds the full KV; the capacity tier does not arise on this workload \\
FastDecode & multi-node$\rightarrow$single-node & against & the unified single-box platform is the constant across all systems \\
\bottomrule
\end{tabular}
\end{table*}

\begin{table*}[t]
\centering
\caption{Reproduction validation anchors. All entries are ratios or hardware ceilings, not cross-hardware absolutes.}
\label{tab:repro-validation}
\footnotesize
\begin{tabular}{@{}lllp{0.42\textwidth}@{}}
\toprule
Baseline & Anchor & Original (reported) & Reproduction (measured) \\
\midrule
SolidAttention & I/O bandwidth ceiling & 7.5~GB/s (SSD) & 25.4~GB/s (DRAM, $=$ H2D peak) \\
ScoutAttention & vs.\ full-KV attention, 8K & 1.78$\times$ (Qwen3-8B) & 1.26$\times$ e2e (same magnitude once the MoE expert share is accounted for) \\
HybridGen & vs.\ MoE-Lightning & 1.87--3.02$\times$ (OPT-13B) & 1.32--3.14$\times$ (Mixtral, native reference) \\
FastDecode & vs.\ MoE-Lightning & --- & 1.39--2.94$\times$ (Mixtral, native reference) \\
HGCA & paradigm equivalence & --- & non-AF ScoutAttention arm $=$ HGCA design (static split, sparsified CPU remainder, cross-engine merge) \\
\bottomrule
\end{tabular}
\end{table*}
\paragraph{Reproduction validation.}
Hardware differences across the original papers are not part of any baseline's mechanism, so validation uses hardware-normalized anchors---bandwidth ceilings and ratios against common references---rather than cross-hardware absolute numbers (Table~\ref{tab:repro-validation}).
SolidAttention's paradigm is I/O-bound: the original's 7.5~GB/s is the SSD ceiling, and our reproduction saturates the DRAM path at the measured H2D peak (25.4~GB/s)---the mechanism reaches the physical limit of its new medium.
ScoutAttention's original paper reports 1.78$\times$ over Transformer full-KV attention (Qwen3-8B, 8K); our reproduction (DS2, 8K, bs=4, retaining 80\% of KV) measures 1.26$\times$ end-to-end against the InplaceKVCache arm (Fig.~\ref{figure10})---the same magnitude once the MoE expert share (experts take 223~ms of a decode step, whereas the original's dense models are attention-dominated) is accounted for.
HybridGen's original paper reports 1.87--3.02$\times$ over MoE-Lightning (OPT-13B, bs=4--16); on Mixtral, our HybridGen reproduction reaches 1.32--3.14$\times$ and our FastDecode reproduction 1.39--2.94$\times$ over natively running MoE-Lightning---the same magnitude against a reference that itself runs without reproduction.

\paragraph{SolidAttention.}
Its original mechanism is SSD-backed dynamic sparse attention: KV merged into coarse-grained blocks, speculative prefetching, fine-grained I/O-compute orchestration.
We replace the storage medium from SSD to DRAM \emph{[favors baseline]}, preserving block-level streaming and sparse selection, with 80\% of KV retained (the quality-retention point reported in the original paper).
SSD$\rightarrow$DRAM removes the I/O bandwidth bottleneck, placing its PCIe streaming bandwidth on par with our ping-pong region at the measured peak (25.4~GB/s); had SSD been retained, the baseline would perform worse.

\paragraph{ScoutAttention.}
Its original mechanism is GPU-resident top-$k$ blocks and CPU layer-ahead precomputation of non-resident blocks (i.e., AF overlap) with asynchronous periodic recall.
We preserve the mechanism and retain 80\% of KV \emph{[mixed]}.
Removing AF overlap recovers the HGCA paradigm~\cite{hgca} (static split, sparsified CPU remainder, cross-engine merge); the non-AF arm in Fig.~\ref{figure9} therefore doubles as the HGCA representative---neither work is open-source, but ScoutAttention's own pipeline diagram confirms the equivalence at matching sparsity.

\paragraph{HybridGen.}
Its original mechanism is CPU-decoupled logits computation, predictive next-layer attention precomputation based on adjacent-layer input similarity (AF overlap), and CXL tiered memory (NUMA-aware).
We preserve logits splitting (CPU computes $qK^{\top}$, reads K but not V, then transfers to GPU for aggregation) and predictive precomputation; the CXL tier is unavailable on our platform \emph{[neutral]}---502~GB of host DRAM already holds the full KV, so the tier's capacity role does not arise on this workload.

\paragraph{FastDecode.}
Its original mechanism is heterogeneous attention execution across multiple CPU nodes.
We shrink to a single node (the GPU does not retain KV; 100\% offloaded to CPU) \emph{[against baseline]}: node-level parallelism is lost, so the original's expected performance is correspondingly higher; the unified single-box platform is the constant across all systems.

\paragraph{Other systems not included.}
In the dense-model regime, SolidAttention has already been compared against FlexGen~\cite{flexgen} and llama.cpp~\cite{llamacpp}; ScoutAttention against HGCA~\cite{hgca} and InfiniGen~\cite{InfiniGen}; HybridGen against FlexGen, Keyformer~\cite{keyformer}, StreamingLLM~\cite{streamingllm}, MoE-Lightning~\cite{Moelightning}, H2O~\cite{h2o}, and InfiniGen.
These systems natively lack the MoE expert execution path required for our workload, and their attention mechanisms are covered by the four baselines above---HGCA in particular is directly represented by the non-AF ScoutAttention arm (see above).
MoE-Lightning (which supports Mixtral and is open-source) is taken as an example: its native design has attention on CPU and experts loaded over PCIe to GPU.
Natively running MoE-Lightning reaches only 34\%--72\% of our reproduced FastDecode on Mixtral: loading one expert over PCIe (14~ms on A100) is much more expensive than the AVX-512-optimized local CPU expert computation (average 4.6~ms at bs=32), and our runtime's expert side universally employs CPU-GPU hierarchical scheduling, widening the gap with pure GPU expert processing.

\paragraph{KV transport layers (LMCache/NIXL).}
These are storage-and-movement layers between inference engines and devices, with no decode path of their own: retrieved KV returns to the engine's GPU for attention (\S\ref{sec:related}).
Turning one into an end-to-end baseline means building a new inference engine around it, and the resulting comparison would again reduce to the attention mechanisms already covered by our four reproduced baselines.

\paragraph{V100 bf16 execution.}
V100 (sm70) has no native bf16 support---Tensor Cores support only fp16, and CUDA cores have no bf16 instructions.
On the GPU side, our system runs bf16 via the PyTorch software promotion path~\cite{pytorch2} (bf16$\rightarrow$fp32) at fp32-level performance without Tensor Core acceleration; on the CPU side, AVX-512-F widens bf16 to fp32 registers for FMA (our Xeon 5318Y lacks AVX-512 BF16).
Since computation is dominated by the CPU and the GPU only handles a small streaming/resident fraction, V100's bf16 deficiency has limited impact on our architecture---this also demonstrates that the CPU-dominated hybrid architecture does not rely on GPU bf16 Tensor Cores.
The three inference frameworks (vLLM, SGLang, and KTransformers), whose kernels are compiled for sm80+, have no corresponding cubin on Volta and thus cannot run on V100.
\end{document}